\documentclass[showpacs,floatfix,superscriptaddres, twocolumn]{revtex4-1}
\usepackage{graphicx,psfrag,amsmath,amssymb,amsfonts,latexsym,color,epsf,graphpap,dcolumn} 
\usepackage{graphicx}
\usepackage{caption}
\usepackage{subcaption}
\usepackage{bbm}
\usepackage{amssymb}
\usepackage{epstopdf}
\usepackage{color, amsmath, bm}
\usepackage[dvipsnames]{xcolor}
\usepackage[hidelinks]{hyperref}
\usepackage{natbib}
\RequirePackage{amsmath}
\usepackage{amsfonts}   
\usepackage{physics} 
\usepackage{comment}
\usepackage{dirtytalk}

\date{today}
\usepackage[latin1]{inputenc}
\usepackage[T1]{fontenc} 
\usepackage[english]{babel}
\usepackage{bm}       
\usepackage{hyperref}
\pdfoutput=1
\begin{document}
\title{Motion induced excitation and radiation from an atom facing a mirror}
\author{C\'esar D. Fosco $^1$ \footnote{fosco@cab.cnea.gov.ar}}
\author{Fernando C. Lombardo$^2$ \footnote{lombardo@df.uba.ar}}
\author{Francisco D. Mazzitelli$^1$ \footnote{fdmazzi@cab.cnea.gov.ar}}

\affiliation{$^1$ Centro At\'omico Bariloche and Instituto Balseiro,
Comisi\'on Nacional de Energ\'\i a At\'omica, 
R8402AGP Bariloche, Argentina}
\affiliation{$^2$ Departamento de F\'\i sica {\it Juan Jos\'e
 Giambiagi}, FCEyN UBA, Facultad de Ciencias Exactas y Naturales,
 Ciudad Universitaria, Pabell\' on I, 1428 Buenos Aires, Argentina }
\date{\today}
\begin{abstract}
We study quantum dissipative effects due to the non-relativistic, bounded,
	accelerated motion of a single neutral atom in the presence of a
	planar perfect mirror, i.e. a perfect conductor
at all frequencies. We consider a simplified model whereby a moving `scalar atom' is coupled to a quantum real
	scalar field, subjected to  either Dirichlet or Neumann boundary
	conditions on the plane. We use an expansion in powers of the departure 
	of the atom with respect to a static average position, to compute the vacuum
	persistence amplitude, and the resulting vacuum decay probability.
We  evaluate transition amplitudes corresponding to the excitation of the atom plus the emission of a particle, and 
show explicitly that the vacuum decay probabilities match the results obtained by integrating the transition amplitudes over the directions of 
the emitted particle. We also compute the spontaneous emission rate of an oscillating atom that is initially in an excited state.

\end{abstract} 
\maketitle
\section{Introduction}\label{sec:intro}
Quantum vacuum fluctuations are at the origin of many interesting
phenomena. A prominent place among them corresponds to the
forces arising between static neutral objects, originated
in the fluctuations of the electromagnetic (EM) field. These Casimir
forces~\cite{books} have been measured with precision in the last decades,
and there have also been theoretical advances on finding their dependence on
the geometry and composition of the bodies. 

Vacuum fluctuations also affect the decay probability of single atoms, via
spontaneous emission. In the proximity of metallic plates or when the atom is in a cavity,
those decay probabilities may change because of the influence of the
boundary conditions on the properties of those fluctuations.  This kind of
effect has been investigated in the context of cavity
electrodynamics~\cite{cavityqed}.

Qualitatively different effects appear when the system is subjected to time
dependent external conditions; for example,  photon creation when macroscopic media
are accelerated or when, being static, a time-dependence  of their electromagnetic
properties is induced. Moreover (under some conditions) even for a medium which moves at a
constant velocity with respect to a static one, a frictional force, termed
`quantum friction', may arise.
These phenomena, broadly named dynamical Casimir effect (DCE) \cite{reviews},
do also manifest themselves at a microscopic level. For example, the
oscillatory motion of the center of mass of an atom, which is initially in
its ground state, can lead to different excited states of the atom-field
system: one of them corresponds to a final state where the atom itself has
been excited, and a photon has been emitted.  Alternatively, the final
state may contain a pair of photons, with the atom still in its ground state.
The latter is the microscopic counterpart of the photon-creation process due to a moving mirror~\cite{NetoMic}. 
A different mechanism, not involving motion, arises when external driving fields produce a time-dependence in the energy levels~\cite{Law}.

Another interesting line of research, closely related to the present paper, focusses on single atoms near a mirror,  and in relative motion to it. For
example, for a two level atom in its excited state,  an oscillating mirror
induces modifications in the decay probability and in the spectrum of the
emitted photons~\cite{Zoller, Passante}. Oscillating atoms near a perfect
mirror have been considered in a number of situations, most of them using
simplifying assumptions, either about the direction of the emitted
photons~\cite{Dolan,Fulling} or considering a scalar rather than the
EM field \cite{Belen}. For uniform acceleration, the relation
between the excitation probability for a moving atom has been compared with
that of a moving mirror, in discussions of the equivalence principle
\cite{Fulling}.  Imperfect mirrors have also been considered, in
particular, in our previous work~\cite{Fosco:2021aih} on quantum
dissipative effects for an atom moving non-relativistically in the presence
of a graphene sheet. The calculations were performed perturbatively in the
coupling constants that define the imperfect media, estimating the dissipative phenomena, in this context, via the imaginary part of the in-out effective action.
This is a ``global'' observable, namely, it accounts for the {\em total\/} probability of vacuum decay. 

In this paper, and for a similar kind of system, we present a two-fold approach to study quantum dissipation and then check their consistency: we first evaluate the imaginary part of the effective action, and then present a more refined study, by evaluating the probability of photon emission as a function of the direction of the emitted photons. 
The consistency of both approaches  is checked by integrating out that emission probability over all the possible angles, and comparing with the probability of vacuum decay. 
We also consider the case of a moving atom that is initially in an excited state, and analyse the dependence of the probability of spontaneous emission on the acceleration of the atom and its distance to the mirror.
We have done all calculations for a model where a real scalar field couples to a scalar model for an atom, which moves in the presence of a ``perfect'' mirror; namely, one  which imposes either Dirichlet or Neumann boundary conditions. 

This paper is organised as follows: in~\ref{sec:scalar}, we introduce the model and evaluate the imaginary part of its effective action, both for Dirichlet and Neumann boundary conditions, to the second order in the amplitude of motion of the atom.
In Sec. ~\ref{sec:prob}, we evaluate the transition probabilities. We find the total decay probability, and also show its
consistency with the finer description of the elementary processes of photon emission and atom excitation. We also discuss the decay or spontaneous emission process in terms of its corresponding probability. Finally, in \ref{sec:conc}, we summarise our main conclusions.

\section{The effective action}\label{sec:scalar}

It has become common usage in research related to both the static and dynamic
Casimir effects, to begin the analysis of the phenomenon being studied in a
simplified setting, with a real scalar field playing the role of
the full EM  field. In some cases, one can even show that transverse electric
and transverse magnetic modes can be described by scalar fields satisfying
either Dirichlet and Neumann boundary conditions, and that the EM field results may be
obtained as the superposition of those two scalar field theories. 

 In our model, an atom will be coupled to a quantum
real scalar field, and we will assume `perfect' boundary conditions
for the scalar field on the flat boundary, i.e.  Dirichlet or Neumann. The
free action for the  massless scalar field shall be given by 
\begin{equation}
{\mathcal S}_{\rm sc}(\phi) \;=\; \frac{1}{2} \,\int d^4x  \, \partial_\mu
	\phi (x) \partial^\mu \phi (x)   \;,
\end{equation}
and boundary conditions will be assumed to be implemented at the level of the functional
integral over gauge field fluctuations. In our conventions, indices from the middle of the Roman alphabet ($i, j\, \ldots$) are assumed to run from $1$ to $3$, while those of the middle of the Greek alphabet ($\mu,
\nu, \ldots$) run from $0$ to $3$, with \mbox{$x^0 \equiv c t$}. In our
conventions, $c \equiv 1$. The metric tensor is, on the other hand, assumed
to be \mbox{$(g_{\mu\nu}) = {\rm diag}(1,-1,-1, -1)$}. Einstein convention
on the sum over repeated indices is also understood, unless explicitly
stated otherwise.

The classical action for the atom is, on the other hand, assumed to be:
\begin{equation}
	{\mathcal S}_{\rm a}=\;\frac{m}{2}\int dt \, 
	\big( \dot{q}^2 - \Omega^2  q^2 \big) +  g \,
	\int dt \,  q(t) 
\; \phi(t,{\mathbf r}(t)) \;,
\end{equation}
where  $q(t)$ plays the role of the electron's degree of freedom, 
${\mathbf r}(t)$ of the atom's center of mass, $m$ is the electron's mass, and $g$ is the coupling constant 
between the electron and the vacuum field.
We will assume that the motion of the  center of mass is bounded and we will denote by ${\bf r}_0$ 
the (time) averaged position.

Following a similar approach to the one of our previous
work~\cite{Fosco:2021aih}, we first integrate out the internal degree of
freedom $q(t)$, what produces an  ``intermediate'' effective action
${\mathcal S}_{\rm eff}(\phi; {\mathbf r})$, given by the expression: 
\begin{equation}
{\mathcal S}_{\rm eff}(\phi; {\mathbf r}) \;=\; {\mathcal S}_{\rm sc}(\phi) + {\mathcal
S}_{\rm I}^{(a)}(\phi; {\mathbf r})
\end{equation}
with
\begin{equation}\label{eq:sia}
{\mathcal S}_{\rm I}^{(a)}(\phi; {\mathbf r}) =- \frac{g^2}{2}
\int dt \int dt'  \Delta(t-t')  \phi(t,{\mathbf r}(t)) 
\phi(t',{\mathbf r}(t')),
\end{equation}
where: $\Delta(t-t') \,=\, \int \frac{d\nu}{2\pi} e^{-i \nu (t-t')} \,
\widetilde{\Delta}(\nu)$, and 
\begin{equation}
	\widetilde{\Delta}(\nu)  \;=\; \frac{i}{m (\nu^2 - \Omega^2 + i
	\epsilon )} \;.
\end{equation}

Then, the complete effective action of the system, functional of the center
of mass coordinates, $\Gamma[{\mathbf r}(t)]$, is  obtained by including
the  real scalar field fluctuations in the presence of the appropriate
boundary conditions. Namely,
\begin{equation}
e^{i \Gamma[{\mathbf r}(t)]} \;=\; \frac{\int \big[{\mathcal D}\phi\big]_m
\, e^{ i {\mathcal S}_{\rm eff}(\phi; {\mathbf r})}}{\int \big[{\mathcal
D}\phi\big]_m \, e^{ i {\mathcal S}_{\rm eff}(\phi; {\mathbf r}_0 )}}. 
\end{equation}
 The subindex `$m$' for the brackets in the field
integration measure has been included to signal that the integral
is over those field configurations that are compatible with the boundary conditions imposed by the mirrors.  
In our case, and as
already advanced, those conditions will be either Dirichlet or Neumann on the plane
$x^3 = 0$. 
We shall not need to actually perform the full integrals above when we evaluate 
$\Gamma[{\mathbf r}(t)]$ to the first order in ${\mathcal S}_I^{(a)}$. 
Indeed, to that order, we just need the correlator of two fields in the presence of
the mirror, which is just the field propagator in the presence of Dirichlet
or Neumann conditions on a plane.
More explicitly, and denoting the contribution of first order in ${\mathcal S}_I^{(a)}$ by $\Gamma^{(a m)}[{\mathbf r}(t)]$, we see that:
\begin{eqnarray}\label{eq:Gammaam}
	\Gamma^{(am)}[{\mathbf r}(t)] & = & \frac{i e^2}{2} 
	\int dt \int dt'  \Delta(t-t') \nonumber \\ 
	&\times &   \langle \phi(t,{\mathbf
	r}(t)) \phi(t',{\mathbf r}(t'))  \rangle^{(m)}, 
\end{eqnarray}
where the symbol $\langle \ldots \rangle^{(m)}$ denotes the functional
averaging 
\begin{equation}\label{eq:eecor}
\langle \ldots \rangle^{(m)} \;=\; \frac{\int \big[{\mathcal
	D}\phi\big]_m \; \ldots  \; e^{i {\mathcal S}_{\rm sc}(\phi)}}{\int
	\big[{\mathcal D}\phi\big]_m \, e^{ i {\mathcal S}_{\rm sc}(\phi)}}
\;.
\end{equation}


\subsection{Dirichlet mirror}

For a Dirichlet mirror, we can simply use the images method, to write:
\begin{align}
\big\langle \phi(x) \phi(y)\big\rangle_m \; = & G_0(x^0-y^0, x^1-y^1,
	x^2-y^2, x^3-y^3) \nonumber\\
	\;- & G_0(x^0-y^0, x^1-y^1, x^2-y^2, x^3+y^3) \;,
\end{align}
where $G_0$ is the free scalar-field propagator
\begin{eqnarray}
	G_0(x^0-&y^0,& x^1-y^1, x^2-y^2, x^3-y^3) =  \nonumber \\ &&=  \int
	\frac{d^4k}{(2\pi)^4} e^{-i k \cdot (x-y)}  \frac{i}{k^2 + i
	\epsilon} .
\end{eqnarray}
Introducing the explicit form of the Dirichlet propagator above into
(\ref{eq:Gammaam}), and using an entirely analogous approach to the one
of~\cite{Belen} we obtain, after some algebra:
\begin{eqnarray}\label{res:dir}
	{\rm Im} [ \Gamma_{mp}^D] &=&  \frac{g^2}{8 \,\Omega \, m } \,
	\int \frac{d^3p}{(2\pi)^3} \;\frac{1}{p} \\ 
&\times &	\big[  f(-{\mathbf p_\shortparallel}, -p^3, -(p + \Omega)) 
	f({\mathbf p_\shortparallel}, p^3, p + \Omega)  \nonumber \\ 
     &-&  f(-{\mathbf p_\shortparallel}, p^3, -(p + \Omega)) 
	f({\mathbf p_\shortparallel}, p^3, p + \Omega) \big]  \nonumber ,
\end{eqnarray}
where ${\bf p} = ({\bf p}_\shortparallel, p^3)$, $p = \vert {\bf p}\vert $, and 
\begin{equation}
f({\mathbf p},\nu) \,=\,
\int_{-\infty}^{+\infty} dt \,  e^{-i {\mathbf p}\cdot {\mathbf r}(t)} \;
e^{i\nu t} \;.
\end{equation}
We obtain more explicit expressions by expanding in powers of the
departure of the atom from an equilibrium position ${\mathbf r}_0$, which
in our choice of coordinates will be of the form ${\mathbf r}_0 = (0,0,a)$,
$a> 0$. 

To the second order in the departure, which we denote ${\mathbf y}(t)$, and using tildes
to denote time Fourier transforms, we find
\begin{equation}
	{\rm Im}[\Gamma_{mp}^D] \;=\; \frac{1}{2} \,
\int_{-\infty}^{+\infty} \,\frac{d\nu}{2\pi} \;  
	\tilde{y}^i(-\nu) \,\tilde{y}^j(\nu) \; m_{\rm D}^{ij}(\nu) \;.
\end{equation}
Because of the presence of the plate, spatial isotropy is lost and $m_{\rm
D}^{ij}(\nu)$ and not proportional to $\delta^{ij}$. Rather, as a $3
\times 3$ matrix, it has the form:
\begin{equation}
	[m_{\rm D}^{ij}(\nu)] \;=\; \left( \begin{array}{ccc}
		m_\shortparallel(\nu) & 0 & 0 \\
		0 & m_\shortparallel(\nu) & 0  \\
                0 & 0 &	m_\perp(\nu)
	\end{array} \right)
\end{equation}
where the parallel
component is 
\begin{eqnarray}
	m_\shortparallel(\nu) &\;=\;& \frac{\pi g^2}{2 m \, \Omega} \;
	\theta(|\nu| - \Omega) \; \int
	\frac{d^3p}{(2 \pi)^3} \, \frac{{\mathbf p_\shortparallel}^2}{2 p} \, 
	[ 1 - \cos( 2 p^3 a) ] \nonumber \\ &\times & \delta( |\nu| - p - \Omega)  \;,
\end{eqnarray}
and results in:
\begin{eqnarray}
m_\shortparallel(\nu) &=& \; \frac{g^2}{8 \pi m \,\Omega} \; \theta(|\nu| - \Omega)\;
(|\nu| - \Omega)^3 \\  &\times &  \Big\{ \frac{2}{3} + \frac{ \cos( 2  (|\nu| -
\Omega ) a)}{ 2  [(|\nu| - \Omega)a]^2 } 
 - \frac{ \sin( 2  (|\nu| - \Omega) a)}{ 4 [(|\nu| - \Omega)a]^3} \Big\} \nonumber .
\end{eqnarray}
The perpendicular component, on the other hand, is given by:
\begin{eqnarray}
m_\perp(\nu)  & \;=\; & \frac{\pi g^2}{2 m \, \Omega} \; \theta(|\nu| -
	\Omega) \,  \int
\frac{d^3p}{(2 \pi)^3} \, \frac{(p^3)^2}{p} \, 
	[ 1 + \cos( 2 p^3 a) ] \nonumber \\ &\times &  \delta(|\nu| - p - \Omega) \;,
\end{eqnarray}
or,
\begin{eqnarray}
&&m_\perp(\nu) = \; \frac{g^2}{4 \pi m \, \Omega} \; \theta(|\nu| - \Omega)\;
	(|\nu| - \Omega)^3  \Big\{ \frac{1}{3} + \\  && \frac{ \cos( 2  (|\nu| -
\Omega) a)}{ 2  [(|\nu| - \Omega)a]^2 }  
 - \frac{ \sin( 2  (|\nu| - \Omega) a)}{ 4 [(|\nu| - \Omega)a]^3}
+ \frac{ \sin( 2  (|\nu| - \Omega) a)}{ 2[(|\nu| - \Omega)a]}
\Big\} \nonumber .
\end{eqnarray}

\begin{figure}
\begin{center}
\includegraphics[scale=0.4]{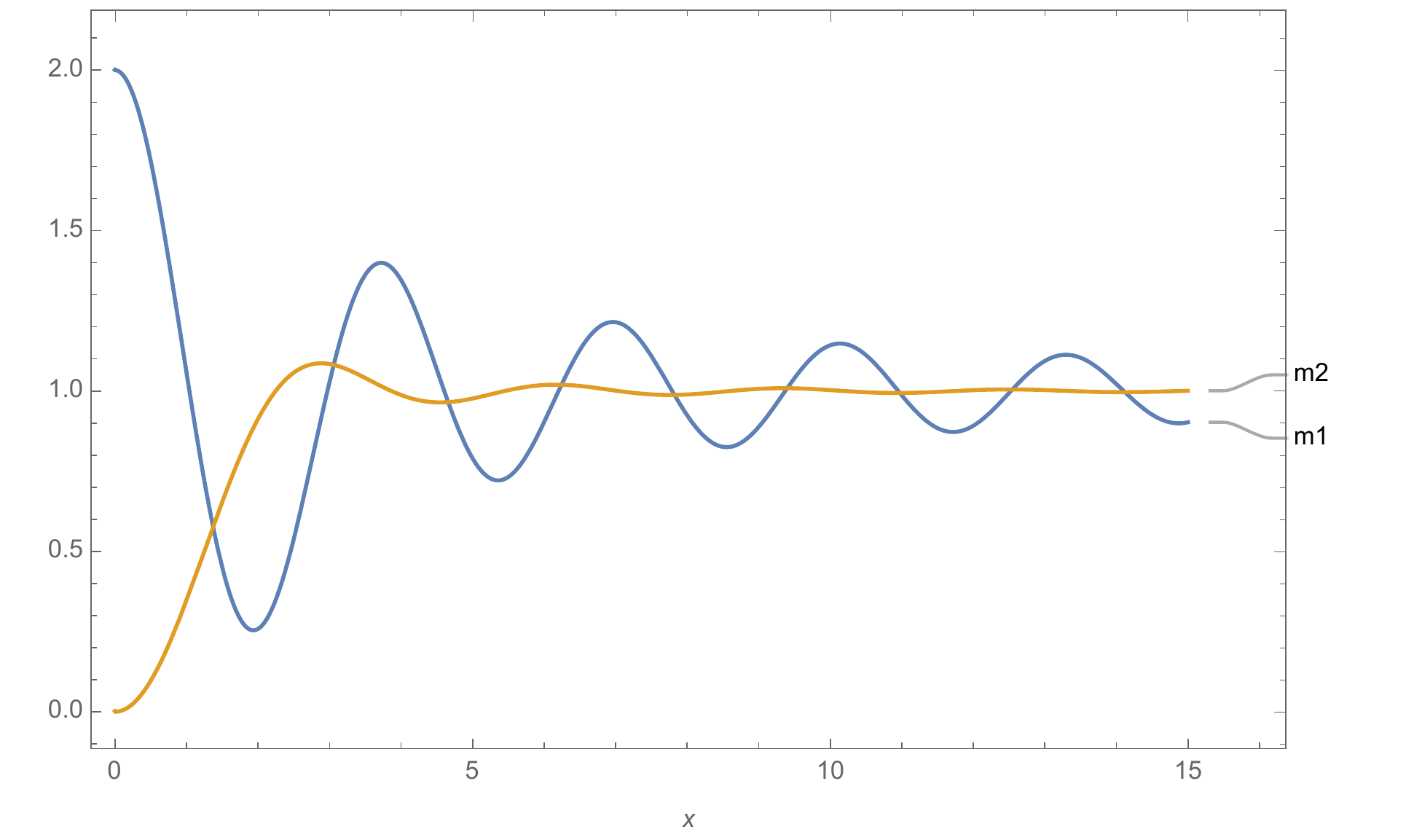}
\caption {Ratios  $m_1 = m_\parallel/m_0$ and  $m_2=  m_\perp/m_0$ as a function of the dimensionless $x = a (\vert \nu\vert - \Omega)$ for Dirichlet boundary conditions on the mirror}
\label{figD}
\end{center}
\end{figure}

It can be seen that both functions, $m_\shortparallel$ and $m_\perp$
approach a common limit $m_0$ far from the plate, i.e., when $|\nu| a \to
\infty$:
\begin{equation}
m_0(\nu) \,=\,\frac{g^2}{12 \pi m\, \Omega } \, \Theta \big( |\nu| - \Omega \big) \;
\big(|\nu| - \Omega \big)^3 \;.
\end{equation}
As expected, this is the result corresponding to an oscillating atom in free space \cite{Belen}. In Fig.\ref{figD} we plot the ratios  $m_\perp/m_0$ and $m_\shortparallel/m_0$ as functions of $a(|\nu|-\Omega)$, where the above mentioned limit can be seen explicitly.
On the contrary, in the opposite regime, i.e., close to the plate, those functions approach rather different limits. Indeed,
$m_\perp$ is enhanced by a factor which approaches $2$, that is $m_\perp(\nu)\approx 2\, m_0 (\nu)$, while $m_\shortparallel$ tends to zero in that limit. These results have a simple interpretation in terms of the images of the oscillating atom in each case. 

\subsection{Neumann mirror}

For a Neumann mirror, one has instead the propagator:
\begin{align}
\big\langle \phi(x) \phi(y)\big\rangle_m \; = & G_0(x^0-y^0, x^1-y^1,
	x^2-y^2, x^3-y^3) \nonumber\\
	\;+ & G_0(x^0-y^0, x^1-y^1, x^2-y^2, x^3+y^3) \;.
\end{align}
Proceeding in an entirely analogous way as for the Dirichlet case,
\begin{equation}
	{\rm Im}[\Gamma_{mp}^N] \;=\; \frac{1}{2} \,
\int_{-\infty}^{+\infty} \,\frac{d\nu}{2\pi} \;  
	\tilde{y}^i(-\nu) \,\tilde{y}^j(\nu) \; m_{\rm N}^{ij}(\nu) \;.
\end{equation}
where now the parallel component is 
\begin{eqnarray}
	m_\shortparallel(\nu) &\;=\; & \frac{\pi g^2}{2 m \, \Omega} \;
	\theta(|\nu| - \Omega) \; \int
	\frac{d^3p}{(2 \pi)^3} \, \frac{{\mathbf p_\shortparallel}^2}{2 p} \, 
	[ 1 + \cos( 2 p^3 a) ] \; \nonumber \\ &\times & \delta( |\nu| - p - \Omega)  \;,
\end{eqnarray}
and results in:
\begin{eqnarray}
m_\shortparallel(\nu) &=& \; \frac{g^2}{4 \pi m \, \Omega} \; \theta(|\nu| - \Omega)\;
	(|\nu| - \Omega)^3 \\ &\times & \Big\{ \frac{1}{3} - \frac{ \cos( 2  (|\nu| -
\Omega) a)}{ 4  [(|\nu| - \Omega)a]^2 } 
	+ \frac{ \sin( 2  (|\nu| - \Omega) a)}{ 8 [(|\nu| - \Omega)a]^3}
\Big\} \nonumber .
\end{eqnarray}
The perpendicular component, on the other hand, is given by:
\begin{eqnarray}
	m_\perp(\nu) &\;=\;& \frac{\pi g^2}{2 m \,\Omega} \; \theta(|\nu| -
	\Omega) \,  \int
\frac{d^3p}{(2 \pi)^3} \, \frac{(p^3)^2}{p} \, 
	[ 1 - \cos( 2 p^3 a) ] \nonumber \\ &\times & \delta(|\nu| - p - \Omega) ,
\end{eqnarray}
or,
\begin{eqnarray}
&&m_\perp(\nu) =  \frac{g^2}{4 \pi m \,\Omega}  \theta(|\nu| - \Omega)
	(|\nu| - \Omega)^3 \, \Big\{ \frac{1}{3} -  \\ && \frac{ \cos( 2  (|\nu| -
\Omega) a)}{ 2  [(|\nu| - \Omega)a]^2 } 
 + \frac{ \sin( 2  (|\nu| - \Omega) a)}{ 4 [(|\nu| - \Omega)a]^3}
- \frac{ \sin( 2  (|\nu| - \Omega) a)}{ 2[(|\nu| - \Omega)a]}
\Big\} \nonumber .
\end{eqnarray}

\begin{figure}
\begin{center}
\includegraphics[scale=0.4]{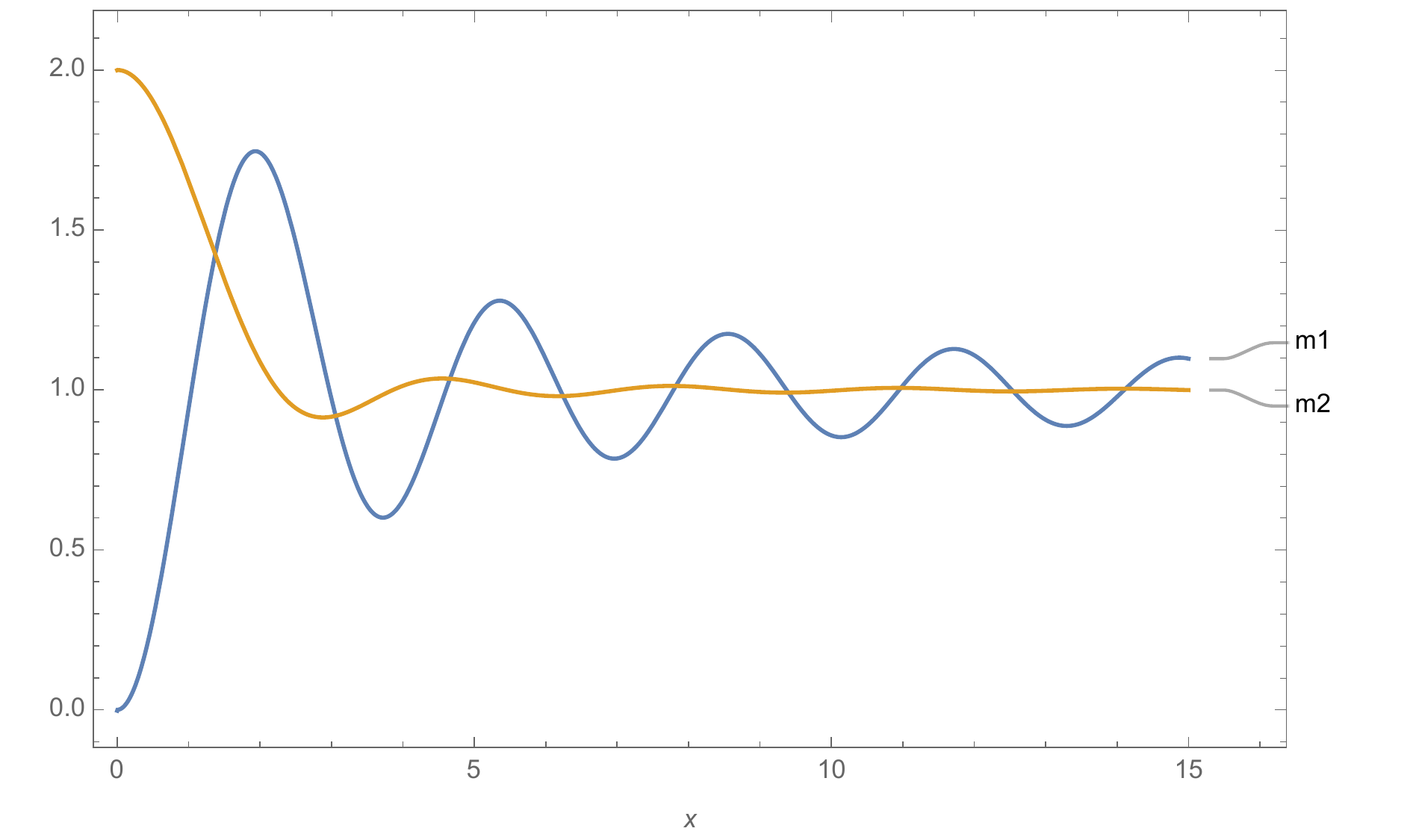}
\caption {Ratios  $m_1 = m_\parallel/m_0$ and  $m_2=  m_\perp/m_0$ as a function of the dimensionless $x = a (\vert \nu\vert - \Omega)$ for Neumann boundary condition on the mirror}
\label{figN}
\end{center}
\end{figure}

Far from the mirror, the results for Neumann boundary conditions also coincide with the free space case. Note, however, that the behaviours of    $m_\shortparallel$ and $m_\perp$ close to the
mirror are reversed, when compared with their Dirichlet counterparts. These results are illustrated in Fig.\ref{figN}.

\section{Transition amplitudes}\label{sec:prob}

The existence of a threshold at $|\nu| = \Omega$, above which there is a
continuum in frequency with a non-vanishing probability of vacuum
decay, suggests that the processes involved correspond to an excitation of
the electron in the atom (hence the threshold $\Omega$), plus the emission
of a massless scalar field particle (the continuum above the threshold).
Besides, we know that the imaginary part of the effective action is related
to the squared modulus of the transition amplitude for the creation of real
particles, when the threshold is reached.
To that end, and having in mind the calculation of the imaginary part of
the effective action to the same order we have used, we consider the
transition matrix $T$, related to the $S$ matrix by
\mbox{$S = I + i T$}, ($I$:
identity operator), to the lowest non-trivial order in the coupling
constant $g$. 

We use standard perturbation theory in the interaction representation,
taking as free Hamiltonian the one corresponding to the free atom plus the
free scalar field (including its boundary conditions). 
Therefore,
\begin{equation}
	S\;=\; {\mathbb T} \exp\left\{ i \,  g \, \int dt \, q(t) \, \phi[t,
	{\mathbf r}(t)] \right\}\;,
\end{equation}
with ${\mathbb T}$ the chronological ordering operator.

The matrix elements of the $T$ matrix between the
initial and final states, to the first order in $g$, are then of the form:
\begin{equation}
	T_{fi} \;\equiv\;  g \, \int dt \, \langle f| 
	q(t) \, \phi[t,{\mathbf r}(t)] | i \rangle \;.
\end{equation}
The operators above evolve independently, according to their respective
free Hamiltonians, thus, for the electron degree of freedom:
\begin{equation}
q(t) \;=\; \frac{1}{\sqrt{2 m\,\Omega}} 
\left( 
	a \, e^{-i \Omega t} 
	\,+\, 
	a^\dagger \, e^{i \Omega t} 
\right)
\end{equation}
where $a$ and $a^\dagger$ denote the standard destruction and
creation operators for the harmonic oscillator. The scalar field, on the
other hand, will have different expansions depending on whether the mirror
imposes Dirichlet or Neumann boundary conditions. We consider these two alternatives below.

\subsection{Dirichlet plane}

In this case, which we consider first,  using the notations: $z \equiv x^3$,
${\mathbf x}_\shortparallel \equiv
(x^1,x^2)$ (and analogously for the components of ${\mathbf k}$), we shall
have:
\begin{equation}
\phi(x)\;=\; \int d^2{\mathbf k}_\shortparallel \,\int_0^\infty  dk_z 
\left[ 
\alpha({\mathbf k}) \,  f_{\mathbf k}(x) 
\,
+
\,
\alpha^\dagger({\mathbf k}) f^*_{\mathbf k}(x) 
\right] \;,
\end{equation}
where 
\begin{equation}
	f_{\mathbf k}(x) \;=\; \frac{1}{\sqrt{4 \pi^3 \omega({\mathbf k})}} \,
	e^{- i \, \omega({\mathbf k}) t + i {\mathbf k}_\shortparallel
	\cdot {\mathbf x}_\shortparallel} \, \sin(k_z z) \;,
\end{equation}
$\omega({\mathbf k}) = |{\mathbf k}|$ and the only non-vanishing commutator
between the operators appearing in the decomposition is
$[\alpha({\mathbf k}) \,,\, \alpha^\dagger({\mathbf
k'})] \,=\, \delta({\mathbf k}-{\mathbf k'})$.

The initial state will be of the form: $|i \rangle \,=\, |0\rangle \otimes |0 \rangle$, where the first factor refers to the 
electron and the second one to the field. 
For this kind of initial state, the only non-vanishing contribution to $T_{fi}$ will be of the form:
\begin{equation}
|f \rangle \;=\; |1\rangle \otimes |{\mathbf k}\rangle \;\;,\;\;\;
|1\rangle \equiv a^\dagger |0\rangle \;,
|{\mathbf k}\rangle \equiv \frac{1}{\mathcal N} \, \alpha^\dagger({\mathbf k})|0\rangle \;.
\end{equation}
The factor $\frac{1}{\mathcal N}$ is included in order to normalize the state $|{\mathbf k}\rangle$, what is needed in order to obtain  
probabilities from $T^{(1)}_{fi}$. If the system is put in a cubic box of length $L$, ${\mathcal N} = \sqrt{\frac{L^3}{(2\pi)^3}}$.

We find:
\begin{eqnarray}
T_{fi} =  \frac{g}{\sqrt{m \Omega \omega({\mathbf k}) L^3}}  \int_{-\infty}^{+\infty} &dt& 
e^{i t (\Omega + \omega({\mathbf k}))} 
e^{- i {\mathbf k}_\shortparallel \cdot {\mathbf r}_\shortparallel(t)} \nonumber \\ &\times &  \sin[k_z r_z(t)]. 
\end{eqnarray}

For parallel motion: $r_z(t) \equiv a$, and to the lowest non-trivial order in the departure \mbox{${\mathbf y}_\shortparallel(t) =
{\mathbf r}_\shortparallel(t)$}, 
\begin{equation}
T_{fi} \;\simeq\; - i \; \frac{g}{\sqrt{m\, \Omega \, |{\mathbf k}| L^3}}
 \; \sin(k_z a) \;
 {\mathbf k}_\shortparallel \cdot
 \tilde{\mathbf y}_\shortparallel(\Omega + |{\mathbf k}|) 
\;. 
\end{equation}
Thus, the probability for the process, with the final particle in an infinitesimal volume in momentum space, $dP_{fi}$, is obtained by 
multiplying $d^3{\mathbf k} \, |T_{fi}^{(1)}|^2$ by the density of final states: $\frac{1}{2}\frac{L^3}{(2\pi)^3}$. Note the factor of $\frac{1}{2}$ with respect to the 
free space case, due to the fact that states with the third component of the momentum reversed are now identical. 

\begin{figure}
    \centering
     \begin{subfigure}[b]{0.27\textwidth}
        \centering
        \includegraphics[width=\linewidth,trim ={.5cm 0 .5cm .5cm}]{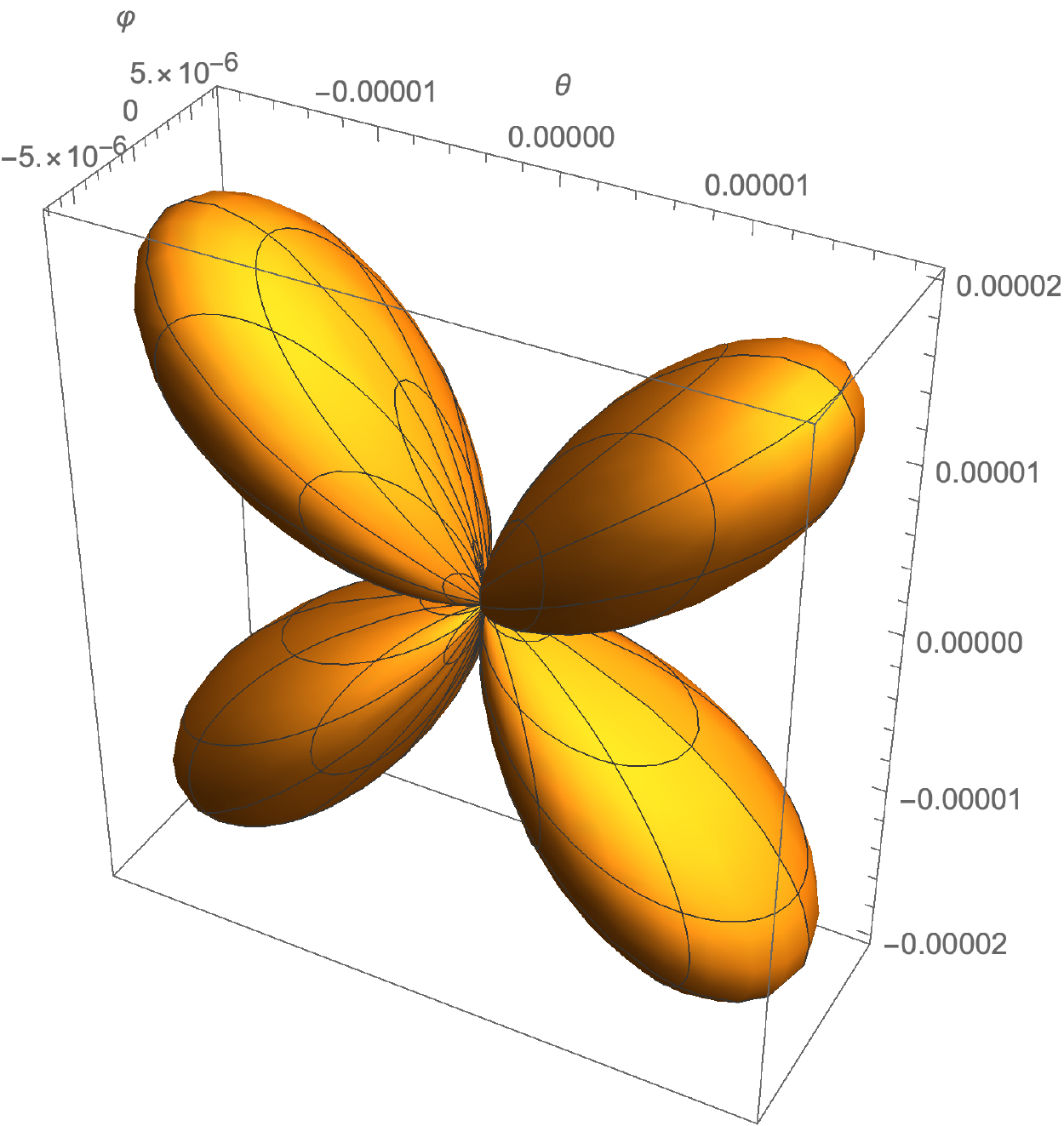}
        \caption{$ k a = 0.001$}
    \end{subfigure}
    \hfill
    \begin{subfigure}[b]{0.25\textwidth}
        \centering
        \includegraphics[width=\linewidth,trim ={.5cm  0 .5cm .5cm}]{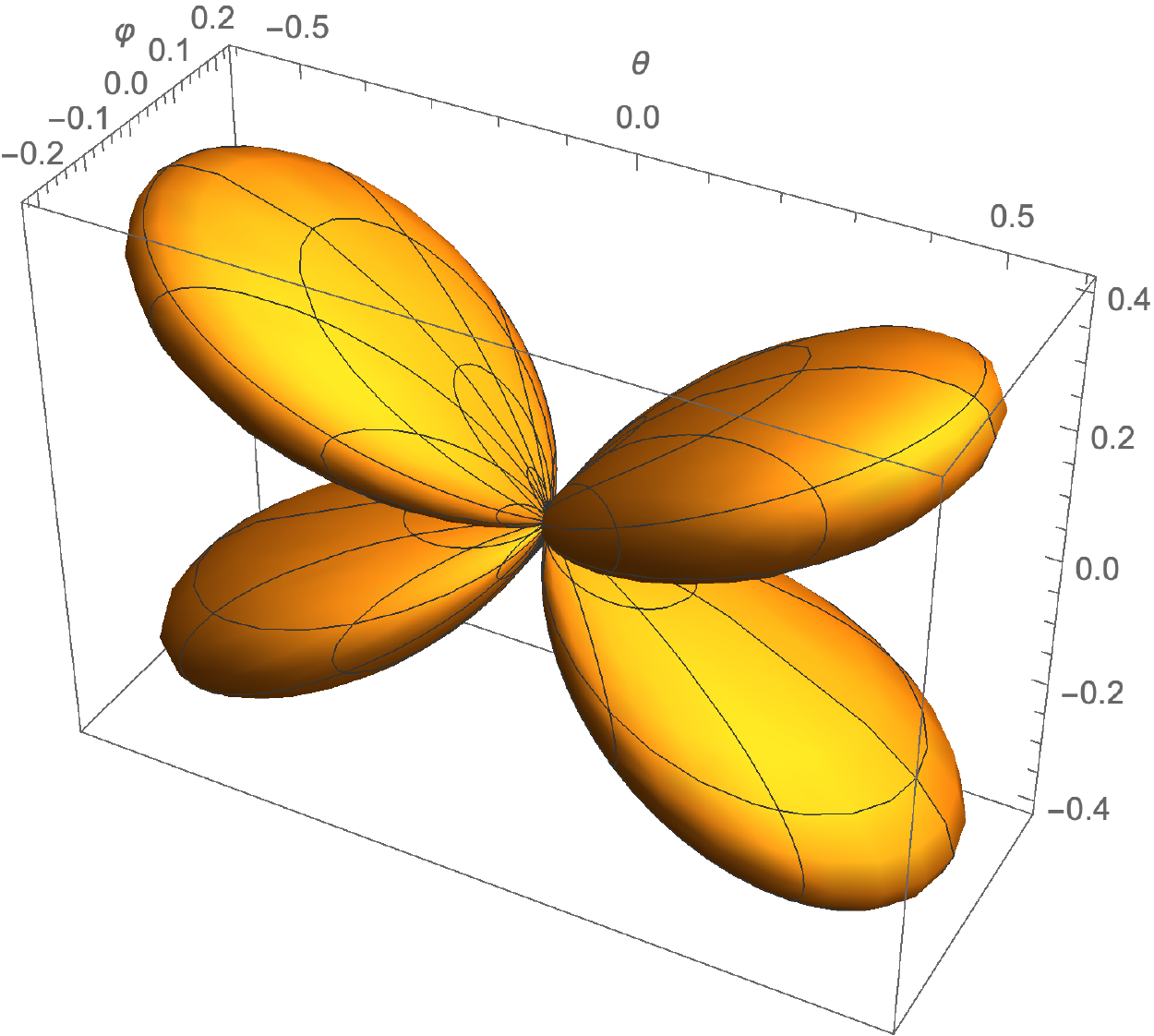}
        \caption{$ k a = 2.5$}
\end{subfigure}
\hfill
\begin{subfigure}[b]{0.36\textwidth}
        \centering
        \includegraphics[width=\linewidth,trim ={.5cm  0 .5cm .5cm}]{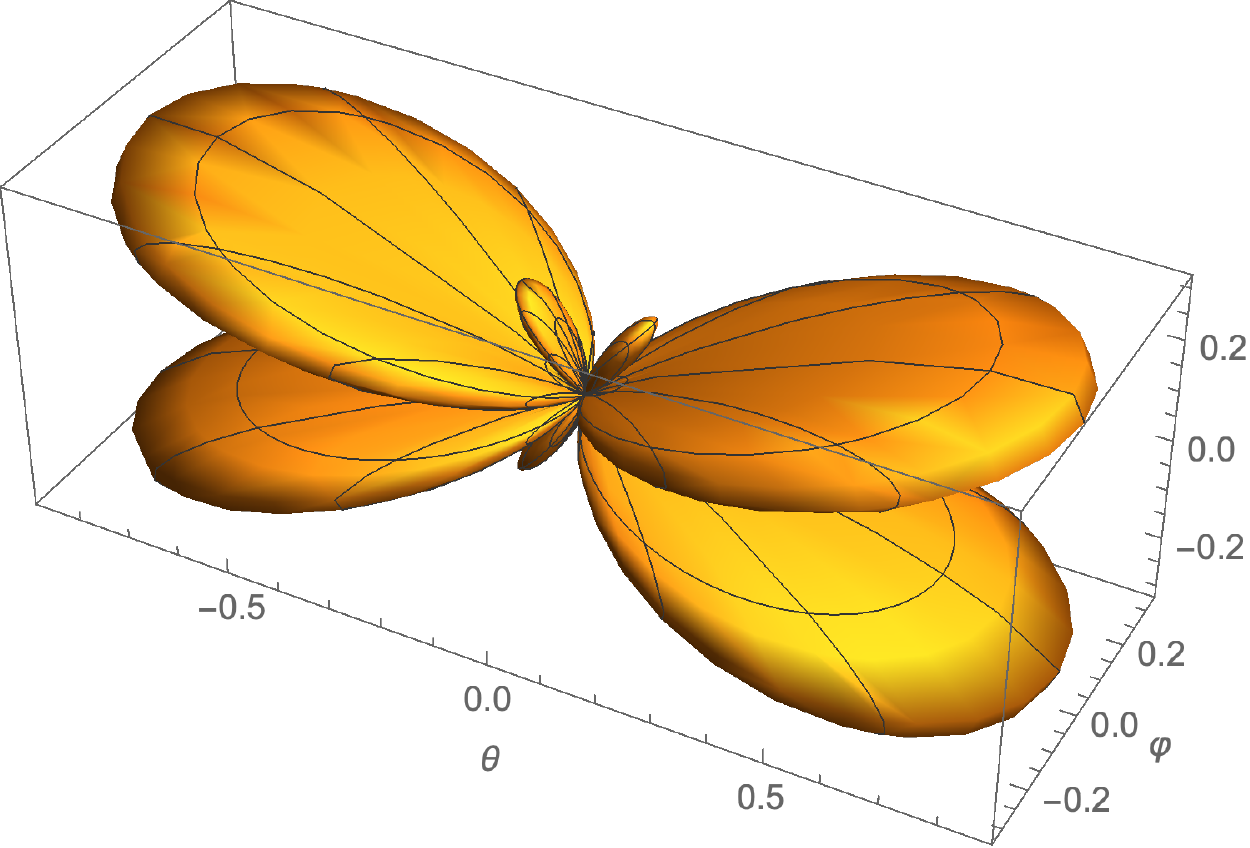}
        \caption{$ k a = 5$}
\end{subfigure}
\caption{We plot here $p_\shortparallel^D$ from Eq.(\ref{Dparaleloang}) as a function of spherical angles $\theta$ and $\varphi$, for Dirichlet boundary 
condition and parallel motion}
\label{Dparalelo}
\end{figure}

Then 
\begin{eqnarray}
dP_{fi}\;=\; \frac{d^3{\mathbf k}}{(2\pi)^3} \,
&\frac{g^2}{2 m \, \Omega \,k}& \, \sin^2(k_z  a) \; 
k^a k^b \\ &\times &  \widetilde{y}^a(-(\Omega + k) ) 
\widetilde{y}^b(\Omega + k ) \nonumber ,
\end{eqnarray}
where $a, b = 1, 2$. 
It is convenient to introduce spherical coordinates centered at the mean position of the atom, such that 
$\tilde{\mathbf y}$ is along the $x^1$ axis. With this choice,  the spatial dependence of the probability is explicitly given by
\begin{eqnarray}
&&dP_{fi} = \frac{g^2}{2 \,(2\pi)^3 \,m\, \Omega } \,k^3\, 
\;\sin^3\theta \, \sin^2(k a \cos\theta) 
\,\cos^2\varphi  \nonumber \\ &&\times  \; |\widetilde{\mathbf y}_\shortparallel(\Omega + k )|^2 \, dk\, 
d\theta \, d\varphi \\
\equiv && \frac{g^2}{2 \,(2\pi)^3 \,m\, \Omega }
p_\shortparallel^D(ka,\theta,\varphi) \; |\widetilde{\mathbf y}_\shortparallel(\Omega + k )|^2 \,  k^3\, \sin\theta\, dk\, 
d\theta \, d\varphi \nonumber ,
\label{Dparaleloang}
\end{eqnarray}
where $p_\shortparallel^D$ is proportional to the probability per unit solid angle. In Fig. \ref{Dparalelo} we plot this function for different values of
the product $ka$. We see that for $ka \leq O(1)$ the angular dependence corresponds to quadrupole radiation, as expected from the images method; 
indeed, in this regime the retardation between a wave emitted by the atom and the one due to its image can be ignored, and therefore  the parallel oscillating 
dipole behaves as a quadrupole when combined with its image companion.

For intermediate values of $ka$, retardation cannot be ignored, and the angular dependence shows a complex structure of peaks, while at very large 
values $\sin^2(k a \cos\theta)$ averages to $1/2$,  the radiation becomes dipolar and coincides with that of an atom oscillating in free  space.

The total probability of this kind of process $P_{fi}$ is obtained by integrating out over all the possible momenta. Rather than integrating over the 
variables above, it is more convenient to proceed as follows:
\begin{eqnarray}
P &=\;&  \int dP_{fi} \\
&=& \; \int d^3{\mathbf k}  \int_{-\infty}^\infty \, 
d\nu  \, \delta(|\nu| - \Omega - k) \;
\frac{g^2}{2 m\, \Omega \,k} \, \sin^2(k_z  a) \nonumber \\
\times && k^a k^b \; \widetilde{y}^a(\nu ) 
\widetilde{y}^b(-\nu ) \nonumber .
\end{eqnarray}
Interchanging the order of the integrals, the one over ${\mathbf k}$ may then be performed exactly. This produces an expression of the form:
\begin{equation}
P\;=\; \frac{1}{2} \, \int \frac{d\nu}{2\pi} \, \rho_\parallel(\nu) \, 
|\tilde{\mathbf y}(\nu)|^2 \;.
\end{equation}

\begin{eqnarray}
\rho_\shortparallel(\nu) &=& \; \frac{g^2}{4\pi m \,\Omega} \; \theta(|\nu| - \Omega)\;
(|\nu| - \Omega)^3 \\ &\times & \Big\{ \frac{2}{3} + \frac{ \cos( 2  (|\nu| -
\Omega ) a)}{ 2  [(|\nu| - \Omega)a]^2 } 
 - \frac{ \sin( 2  (|\nu| - \Omega) a)}{ 4 [(|\nu| - \Omega)a]^3} \Big\} \nonumber .
\end{eqnarray}
Note that this is in agreement with our results for the imaginary part, since $P$ is twice the imaginary part of the effective action.

\begin{figure}
    \centering
     \begin{subfigure}[b]{0.16\textwidth}
        \centering
        \includegraphics[width=\linewidth,trim ={.5cm 0 .5cm .5cm}]{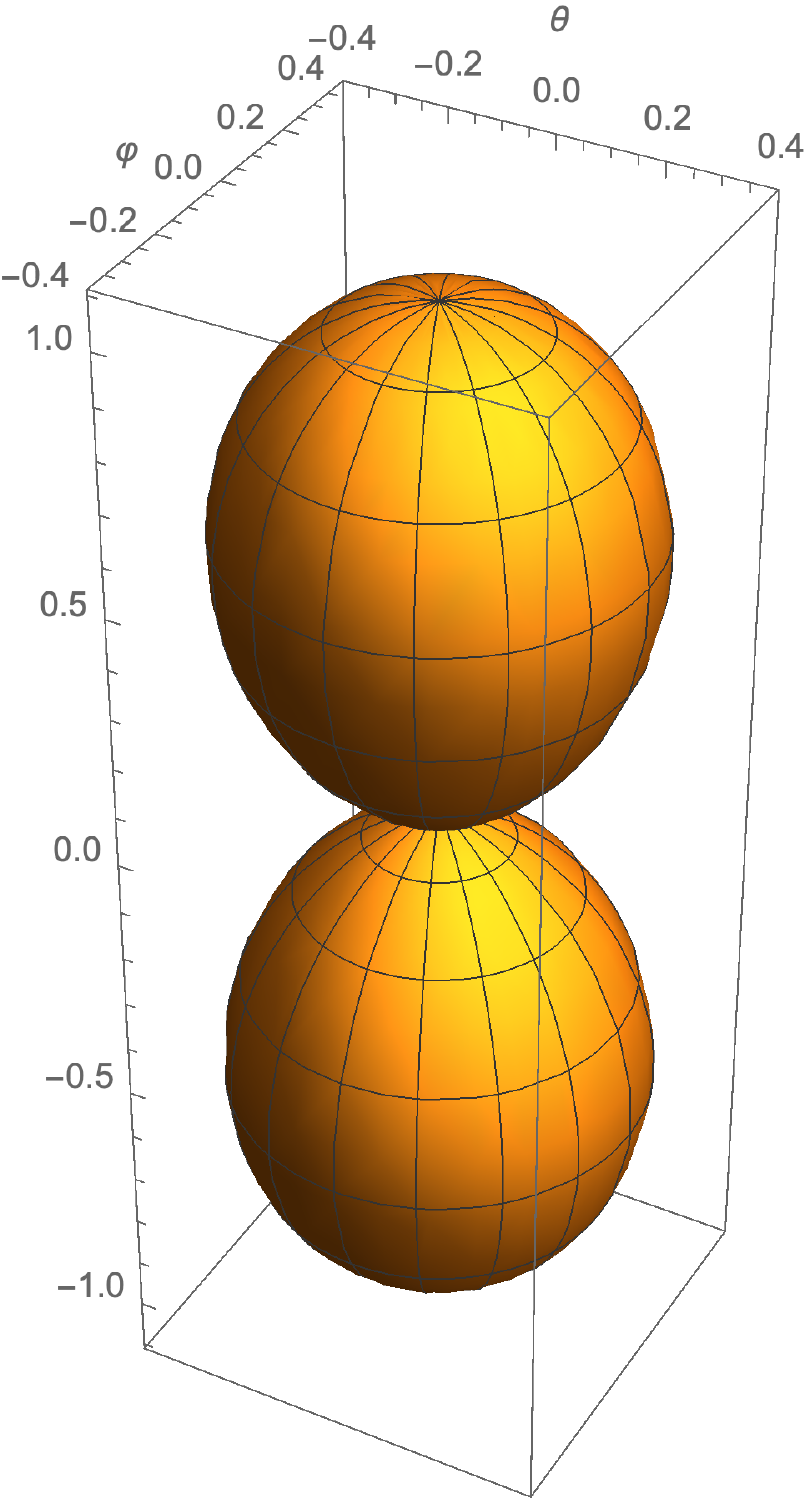}
        \caption{$ k a = 0.001$}
    \end{subfigure}
    \hfill
    \begin{subfigure}[b]{0.2\textwidth}
        \centering
        \includegraphics[width=\linewidth,trim ={.5cm  0 .5cm .5cm}]{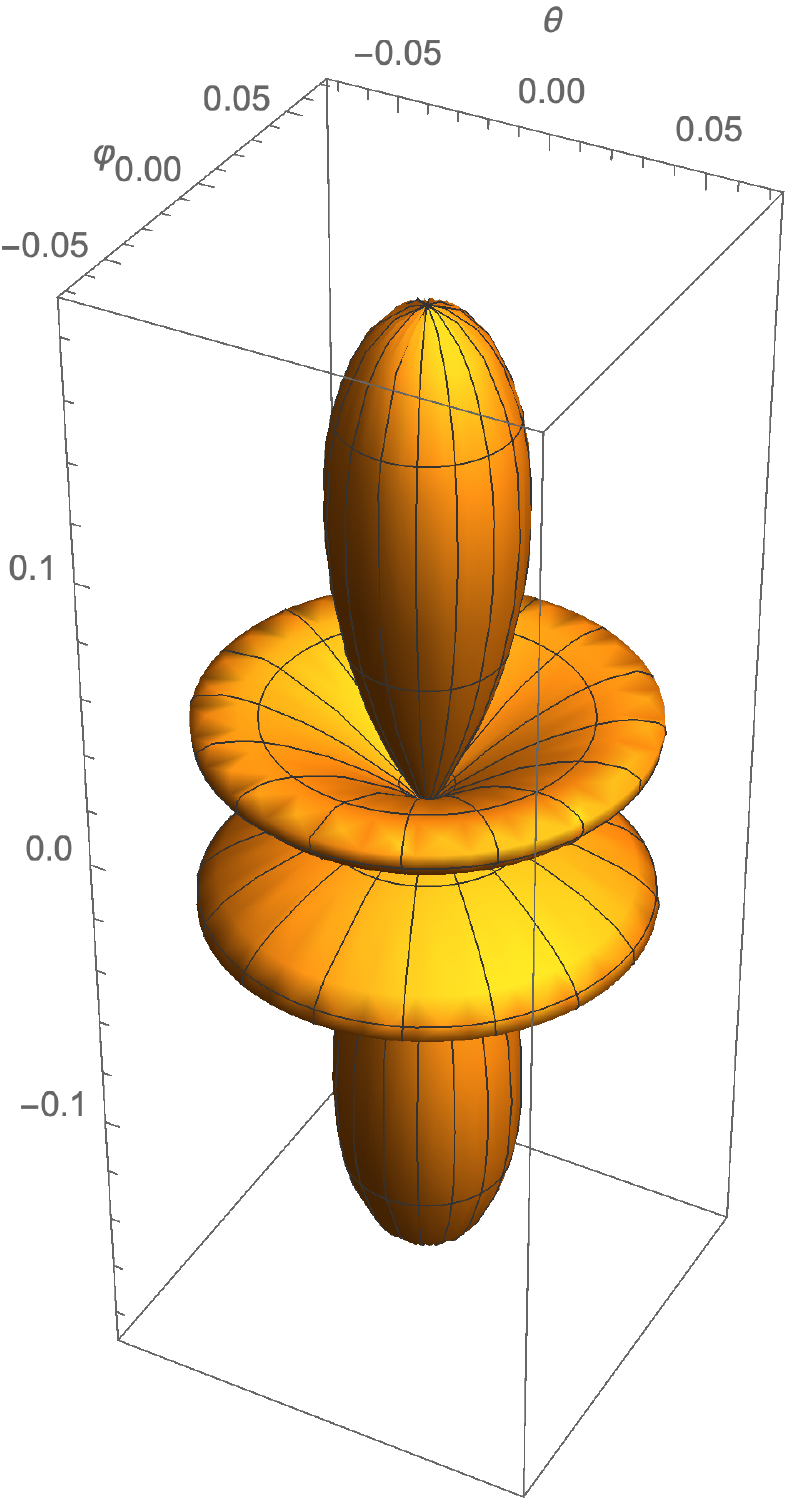}
        \caption{$ k a = 2.5$}
\end{subfigure}
\hfill
\begin{subfigure}[b]{0.24\textwidth}
        \centering
        \includegraphics[width=\linewidth,trim ={.5cm  0 .5cm .5cm}]{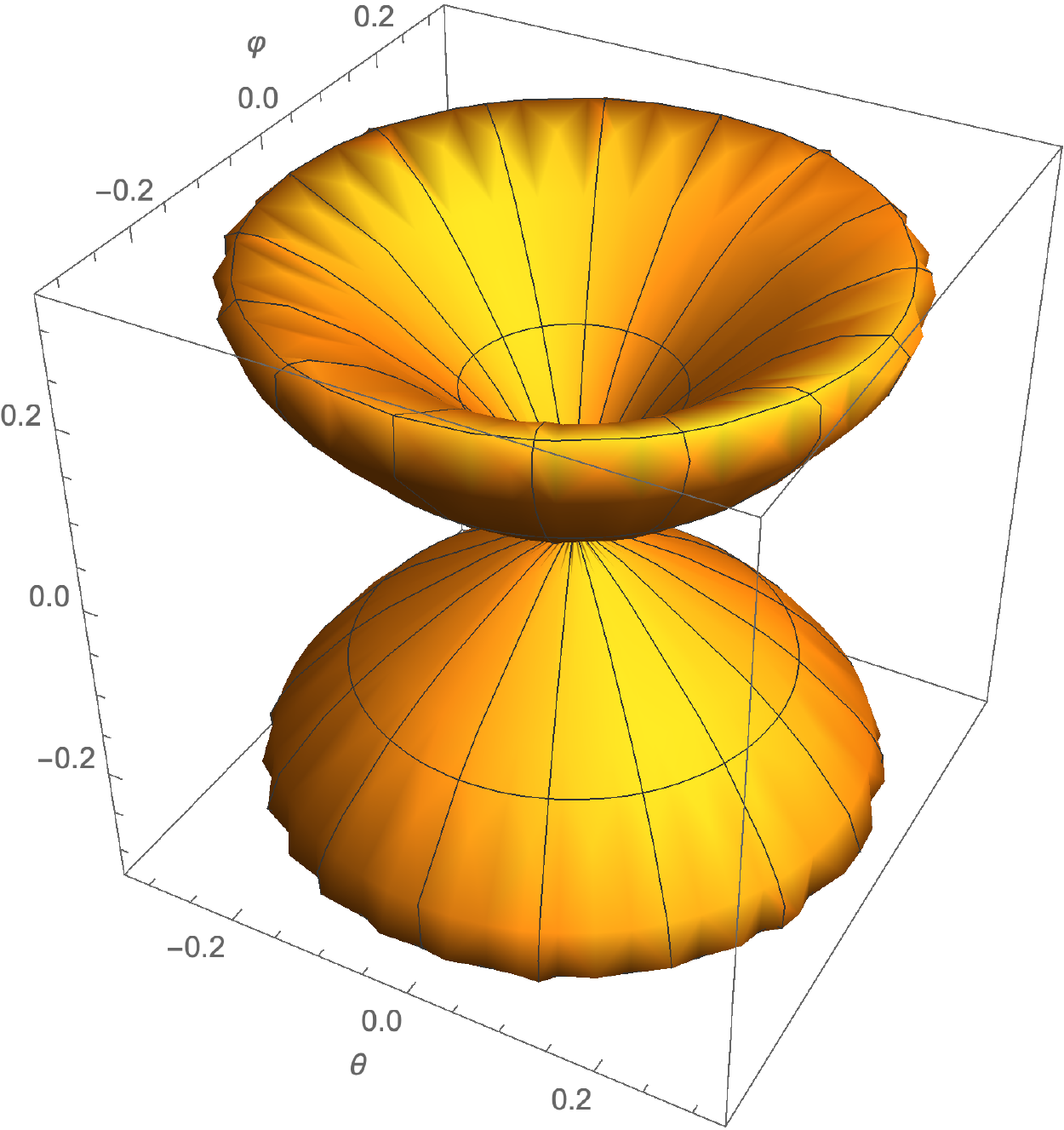}
        \caption{$ k a = 5$}
\end{subfigure}
\caption{We plot here $p_\perp^D$ from Eq.(\ref{Dperpang}) as a function of spherical angles $\theta$ and $\varphi$, for Dirichlet boundary condition and perpendicular motion}
\label{Dperp}
\end{figure}

For motion perpendicular to the plane,
\begin{equation}
T_{fi} \;\simeq\; \frac{g}{\sqrt{m\, \Omega \, |{\mathbf k}|L^3}}
 \; \cos(k_z a) \; k_z \tilde{y}_\perp(\Omega + |{\mathbf k}|) 
\;,
\end{equation}
and
\begin{equation}
dP_{fi}\;=\; \frac{d^3{\mathbf k}}{(2\pi)^3} \,
\frac{g^2}{2 m \, \Omega \,k} \, \cos^2(k_z  a) \; 
k_z^2 \; |\widetilde{y}_\perp(\Omega + k)|^2 \;.
\end{equation}
In this case there is of course axial symmetry with respect to the direction of motion; therefore we may integrate the probability along
$\varphi$ angle
\begin{eqnarray}
dP_{fi}\;&=&\; \frac{g^2}{2 \,(2\pi)^2 \,m\, \Omega} \,k^3\;\sin \theta \, \cos^2\theta \, \cos^2(k a \cos\theta) \nonumber \\ &\times &  |\widetilde{y}_\perp(\Omega + k )|^2 \, dk\, d\theta \\
\;&\equiv &\; \frac{g^2}{2 \,(2\pi)^2 \,m\, \Omega} p_\perp^D (ka,\theta) \; |\widetilde{y}_\perp(\Omega + k )|^2 \, k^3 \sin \theta \, dk\, d\theta\, \nonumber .
\label{Dperpang}
\end{eqnarray}
In Fig. \ref{Dperp} and  we plot the function 
$p_\perp^D (ka,\theta)$ for different values of $ka$. We can see that the radiation has a dipolar pattern both at small and very large distances: at very 
small distances the image dipole {\em reinforces\/} (without retardation and therefore no interference) the effect of the vertical oscillating dipole associated to the 
moving atom, while at large distances we recover the result for the oscillating atom in free space.

Again, the total probability becomes identical to twice the imaginary part of the effective action (evaluated to the second order in $g$).

In the same way, we shall study the spontaneous decay process in which the initial state is now given by $|i \rangle \,=\, |1\rangle \otimes |0 \rangle$ (i.e. the atom in 
the excited state and the field in vacuum) and the final state is given by $
|f \rangle \;=\; |0\rangle \otimes |{\mathbf 1}\rangle$. 
It is straightforward to check that in this case there is no threshold at $\vert \nu \vert = \Omega$ and the result is given for the matrix element $T_{fi}$ in the parallel motion 
is 

\begin{eqnarray}
T_{fi} = \frac{g}{\sqrt{m\, \Omega \, \omega({\mathbf k}) L^3}} \; 
\int_{-\infty}^{+\infty} &dt& \, 
e^{i t (- \Omega + \omega({\mathbf k}))} \, 
e^{- i {\mathbf k}_\shortparallel \cdot {\mathbf r}_\shortparallel(t)} \nonumber \\ &\times & \sin[k_z r_z(t)] ,
\end{eqnarray}
for $r_z(t) \equiv a$, and to the lowest non-trivial order in the departure \mbox{${\mathbf y}_\shortparallel(t) =
{\mathbf r}_\shortparallel(t)$}, 
\begin{eqnarray}
T_{fi} \simeq   \frac{g}{\sqrt{m\, \Omega \, |{\mathbf k}| L^3}}
& \sin[k_z a]&  \Big[ 2\pi \delta(\Omega - \omega(k))  \\  & -& i \;
 {\mathbf k}_\shortparallel \cdot
 \tilde{\mathbf y}_\shortparallel(- \Omega + |{\mathbf k}|) \Big]
\nonumber, 
\end{eqnarray}
where, unlike the excitation case examined previously, there is a first term independent of 
the amplitude of the oscillation $\tilde{\mathbf y}_\shortparallel$, that corresponds to the spontaneous decay
for a static atom in front of a mirror. 
The probability for the decay process is now given 
by 

\begin{eqnarray}
dP_{fi} &=&  \frac{d^3{\mathbf k}}{(2\pi)^3} 
\frac{g^2}{2 m  \Omega k}   \sin^2(k_z  a)  \Big[ 4 \pi^2 T \delta(\Omega  - \vert {\mathbf k}\vert)  \nonumber \\ &+&   
k^a k^b \widetilde{y}^a(-(-\Omega + \vert {\mathbf k}\vert)) 
\widetilde{y}^b(-\Omega + \vert {\mathbf k}\vert ) \Big].
\end{eqnarray}

For the motion perpendicular to the plane we have

 \begin{eqnarray}
T_{fi} &\simeq & \frac{g}{\sqrt{m\, \Omega \, |{\mathbf k}|L^3}}
 \; \Big[ 2\pi \delta(\Omega - \vert {\mathbf k}\vert ) \sin(k_z a)  \nonumber \\ &+&  k_z \tilde{y}_\perp(-\Omega + |{\mathbf k}|) \cos(k_z a) \Big]
\end{eqnarray}
and
\begin{eqnarray}
dP_{fi} &\;=\; & \frac{d^3{\mathbf k}}{(2\pi)^3} \,
\frac{g^2}{2 m \, \Omega \,k}  \; \Big[ 4 \pi^2 T \delta(\Omega - \vert {\mathbf k}\vert )  \sin^2(k_z  a) \nonumber \\ &+&  
k_z^2 \; |\widetilde{y}_\perp(- \Omega + \vert {\mathbf k}\vert)|^2  \cos^2(k_z  a) \Big]\;, 
\end{eqnarray}
assuming that $\tilde{y}_\perp(0) = 0$. 

It is worth to remark that in both situations (parallel o perpendicular motions), if the atom's center of mass oscillates harmonically with 
a frequency  $\Omega_{\rm cm} < \Omega$, the spectrum of the emitted photons will have three different frequencies 
$\omega=\Omega,\, \Omega\pm \Omega_{\rm cm}$. A similar phenomenon occurs for an atom in front of an oscillating mirror
\cite{Zoller,Passante}, in the adiabatic approximation. Only two frequencies would be present in the spectrum when $\Omega_{\rm cm} > \Omega$.

\subsection{Neumann plane}

We now have a different expansion for the field:
\begin{equation}
\phi(x)\;=\; \int d^2{\mathbf k}_\shortparallel \,\int_0^\infty  dk_z 
\left[ 
\alpha({\mathbf k}) \,  g_{\mathbf k}(x) 
\,
+
\,
\alpha^\dagger({\mathbf k}) g^*_{\mathbf k}(x) 
\right] \;,
\end{equation}
where 
\begin{equation}
	g_{\mathbf k}(x) \;=\; \frac{1}{\sqrt{4 \pi^3 |{\mathbf k}|}} \,
	e^{- i \, |{\mathbf k}| t + i {\mathbf k}_\shortparallel
	\cdot {\mathbf x}_\shortparallel} \, \cos(k_z z) \;.
\end{equation}

\begin{figure}
    \centering
     \begin{subfigure}[b]{0.32\textwidth}
        \centering
        \includegraphics[width=\linewidth,trim ={.5cm 0 .5cm .5cm}]{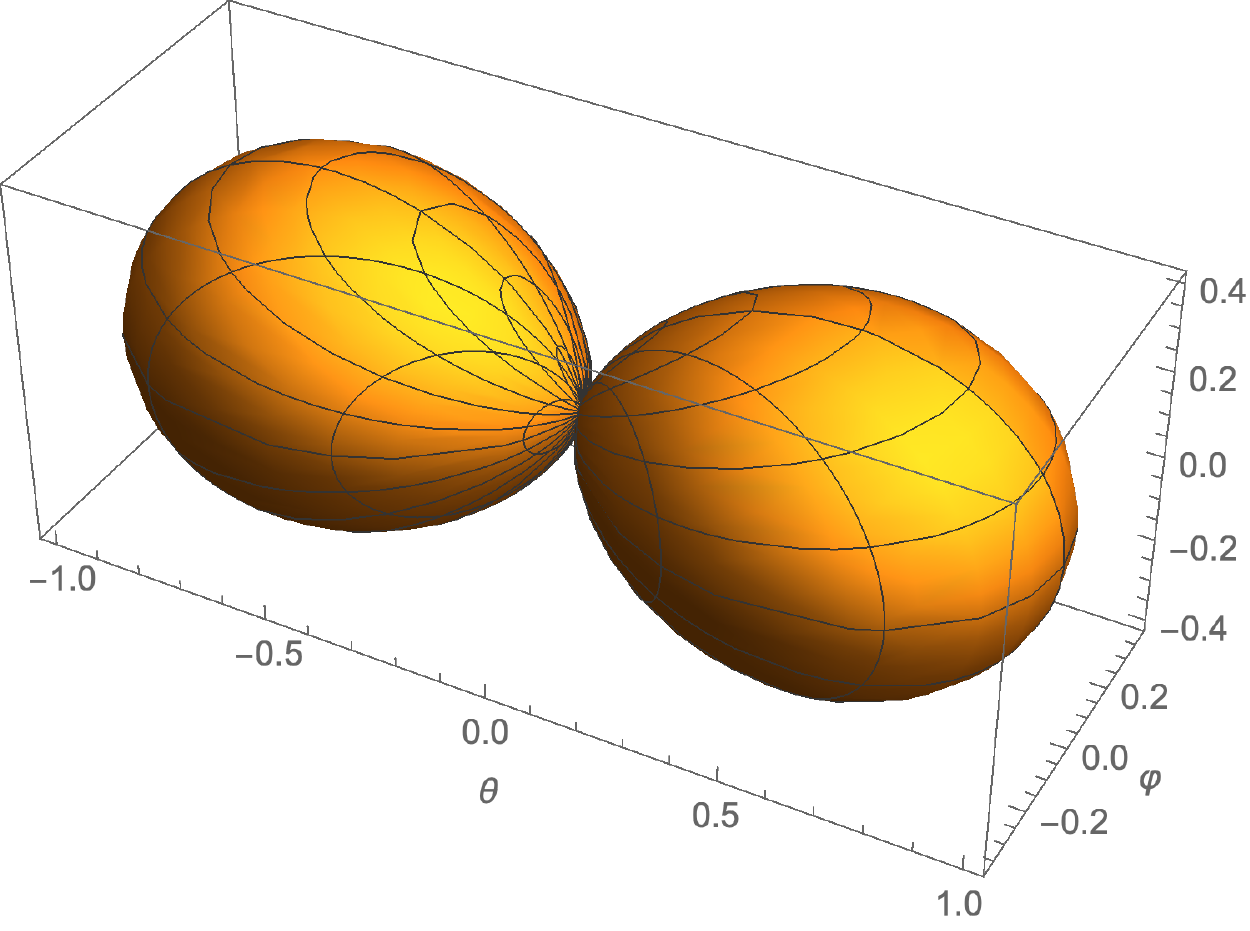}
        \caption{$ k a = 0.001$}
    \end{subfigure}
    \hfill
    \begin{subfigure}[b]{0.32\textwidth}
        \centering
        \includegraphics[width=\linewidth,trim ={.5cm  0 .5cm .5cm}]{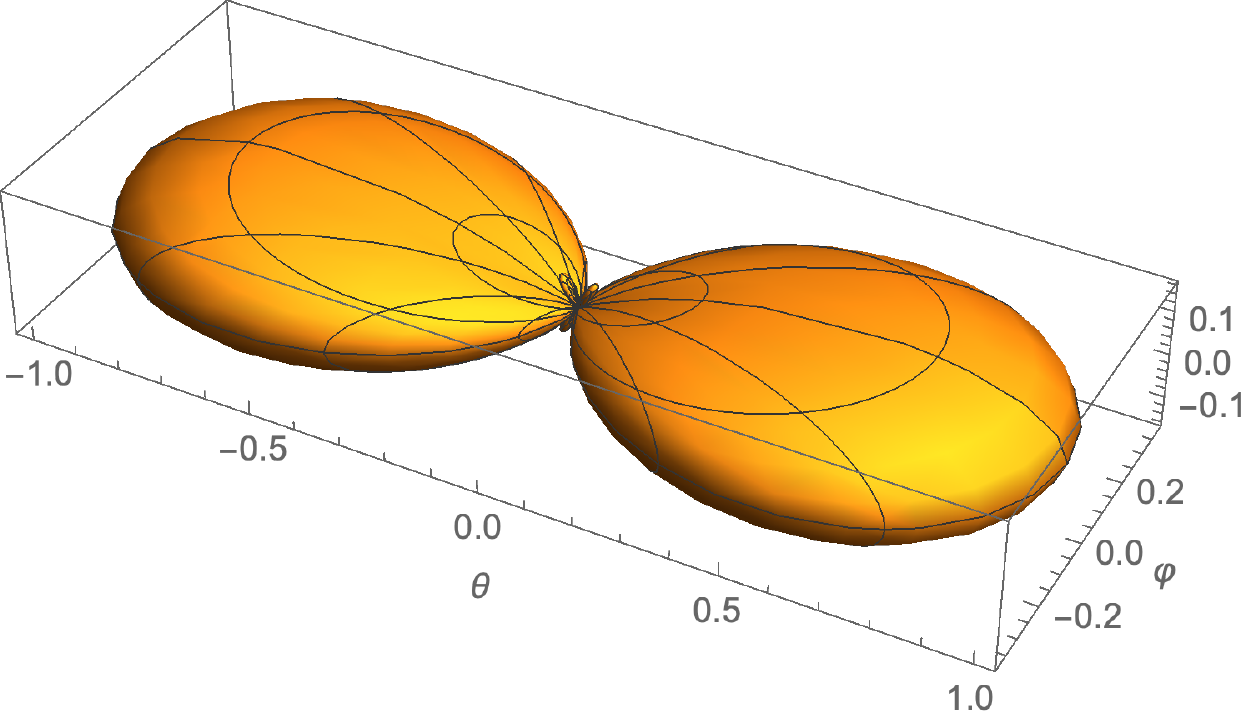}
        \caption{$ k a = 2.5$}
\end{subfigure}
\hfill
\begin{subfigure}[b]{0.32\textwidth}
        \centering
        \includegraphics[width=\linewidth,trim ={.5cm  0 .5cm .5cm}]{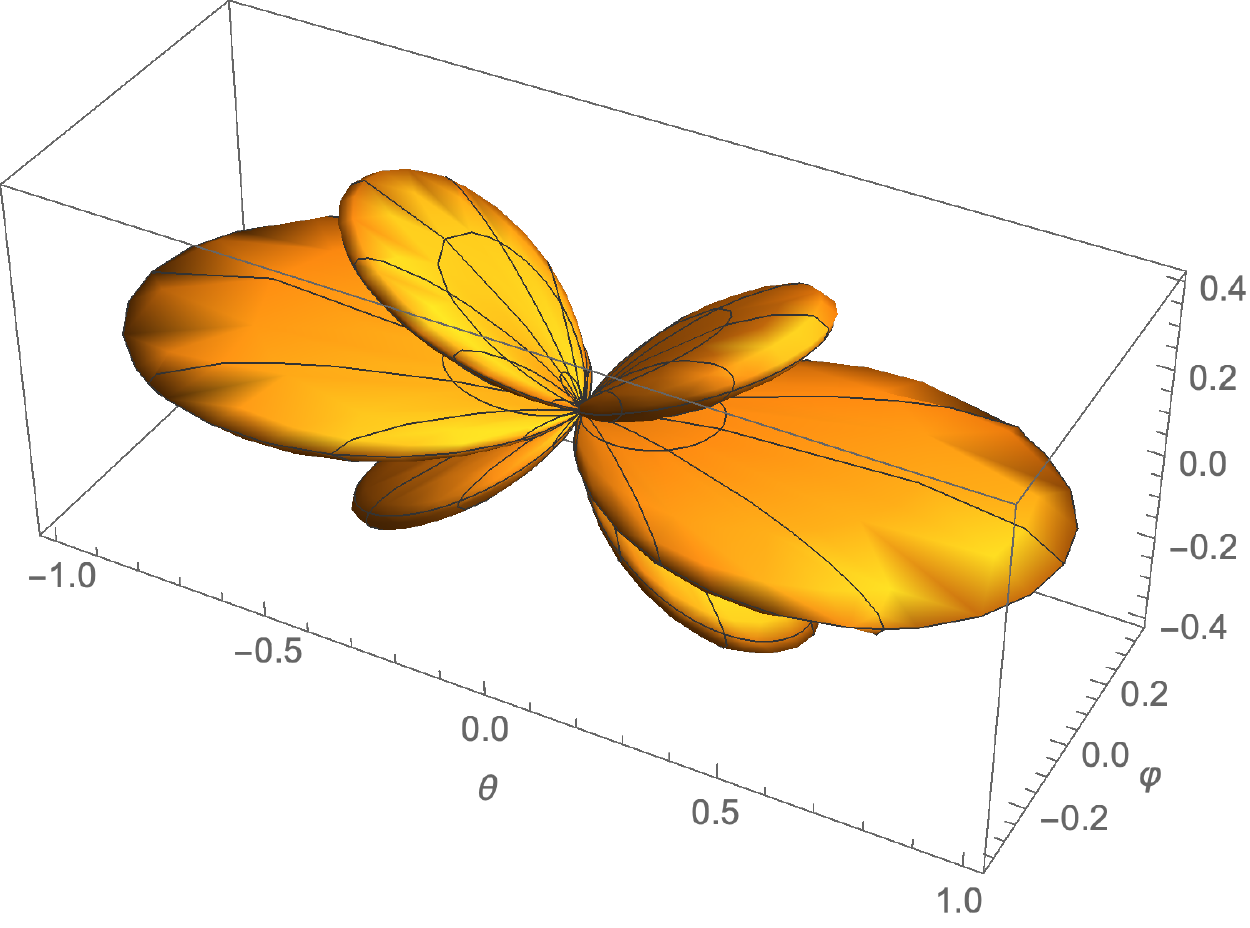}
        \caption{$ k a = 5$}
\end{subfigure}
\caption{We plot here $p^N_\shortparallel$ from Eq.(\ref{Nparaleloang}) as a function of spherical angles $\theta$ and $\varphi$, for Neumann boundary 
condition and parallel motion}
\label{Nparalelo}
\end{figure}

\begin{figure}
    \centering
     \begin{subfigure}[b]{0.14\textwidth}
        \centering
        \includegraphics[width=\linewidth,trim ={.5cm 0 .5cm .5cm}]{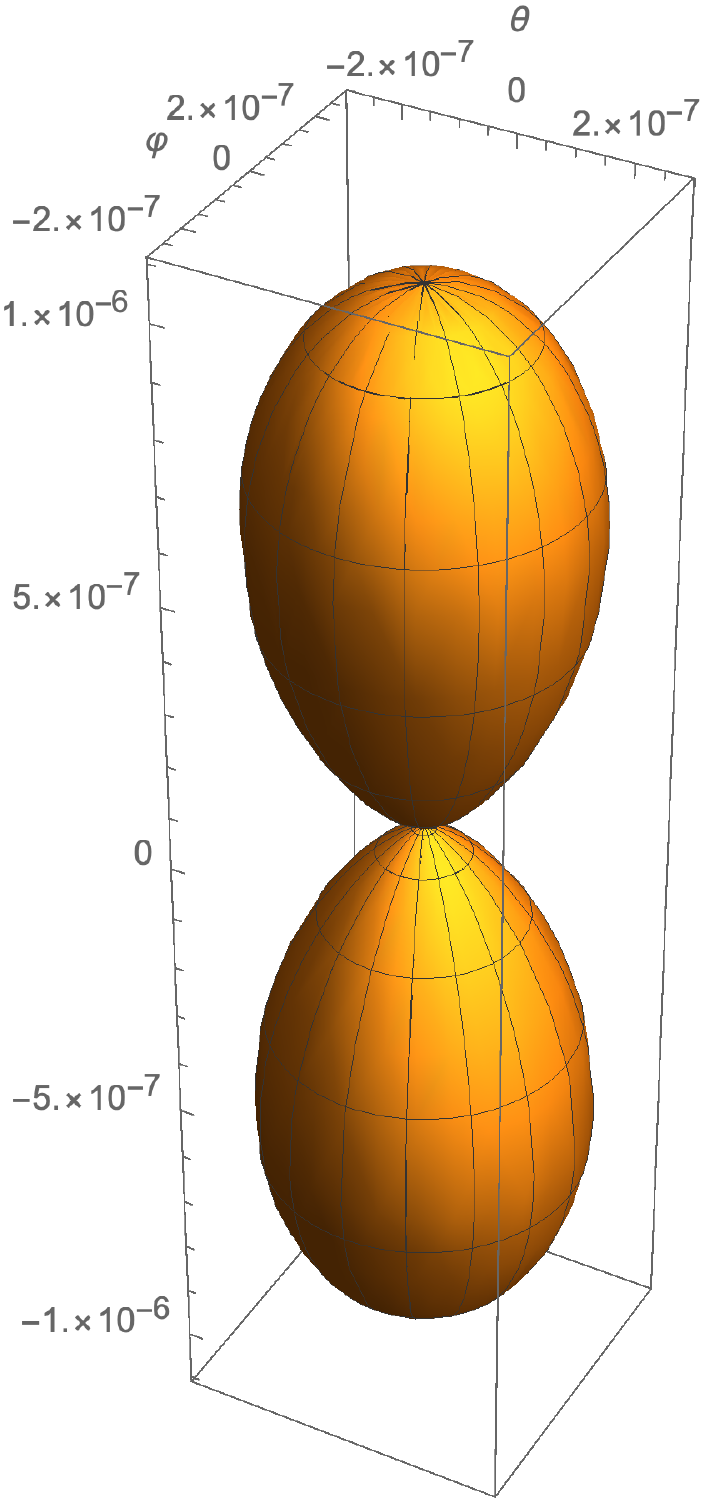}
        \caption{$ k a = 0.001$}
    \end{subfigure}
    \hfill
    \begin{subfigure}[b]{0.24\textwidth}
        \centering
        \includegraphics[width=\linewidth,trim ={.5cm  0 .5cm .5cm}]{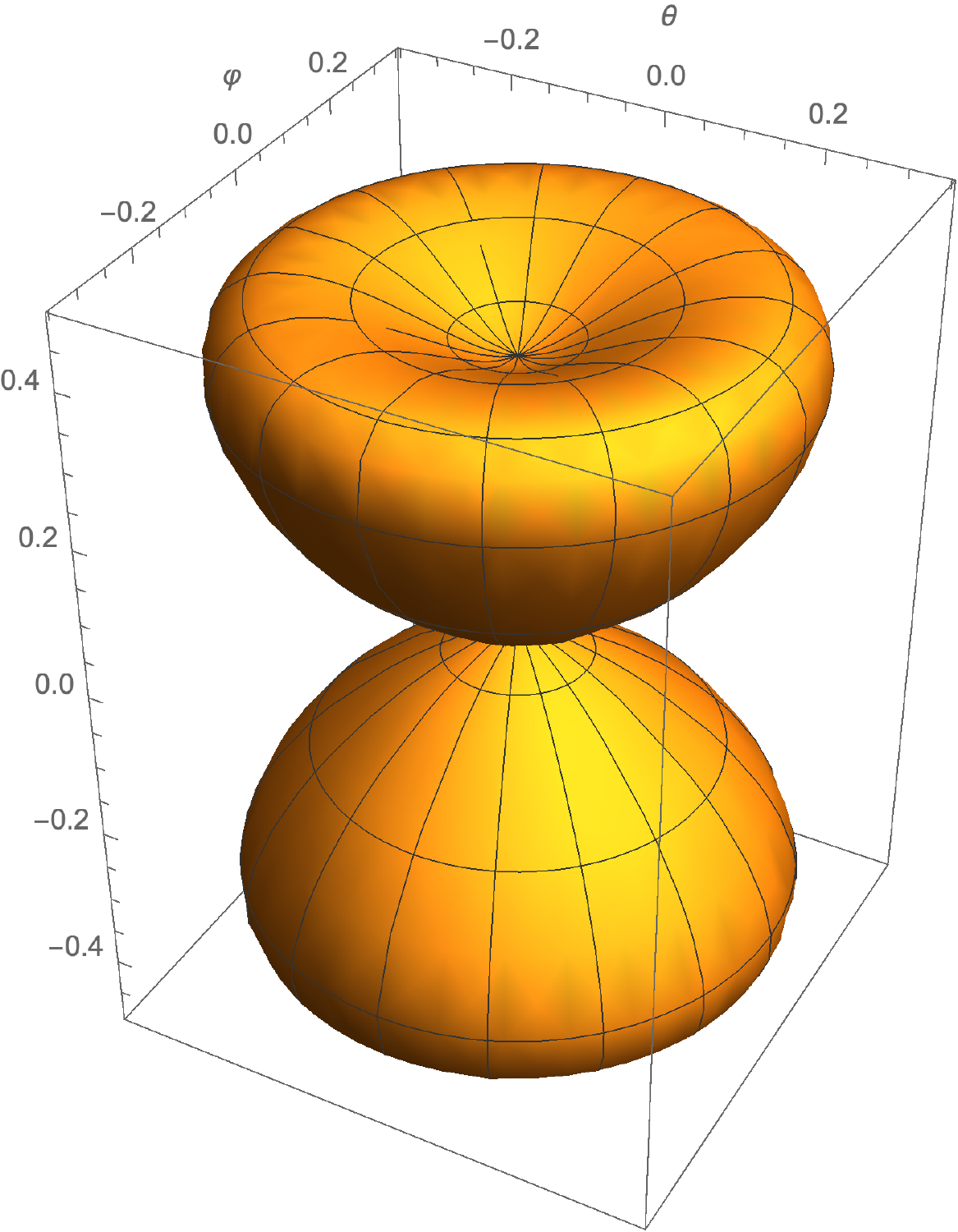}
        \caption{$ k a = 2.5$}
\end{subfigure}
\hfill
\begin{subfigure}[b]{0.16\textwidth}
        \centering
        \includegraphics[width=\linewidth,trim ={.5cm  0 .5cm .5cm}]{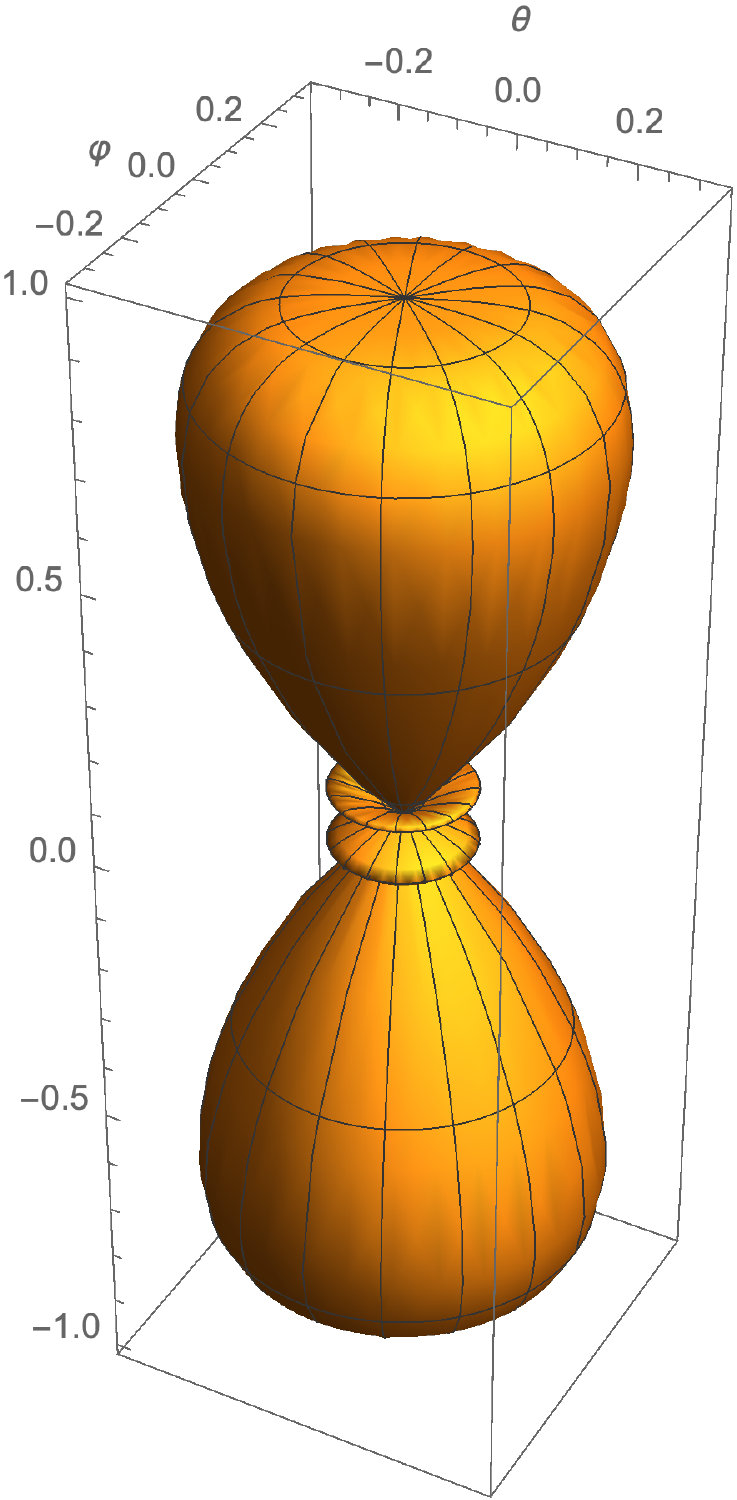}
        \caption{$ k a = 5$}
\end{subfigure}
\caption{We plot here $p_\perp^N$ from Eq.(\ref{Nperpang}) as a function of spherical angles $\theta$ and $\varphi$, for Neumann boundary 
condition and perpendicular motion}
\label{Nperp}
\end{figure}

Since the interaction term is exactly the same as for the Dirichlet case,
it is immediate to find the probabilities in this case. For parallel motion:
\begin{eqnarray}
dP_{fi} &=& \frac{g^2}{2 \,(2\pi)^3 \,m\, \Omega } \,k^3\, 
\;\sin^3\theta \, \cos^2(k a \cos\theta) 
\,\cos^2\varphi  \nonumber \\ &\times &  |\widetilde{\mathbf y}_\shortparallel(\Omega + k )|^2 \, dk\, 
d\theta \, d\varphi \nonumber\\
&\equiv &\; \frac{g^2}{2 \,(2\pi)^3 \,m\, \Omega }  
p_\shortparallel^N(ka,\theta,\varphi)\,
  \; |\widetilde{\mathbf y}_\shortparallel(\Omega + k )|^2 
  k^3 \nonumber \\ &\times &  \sin\theta\, dk\, 
d\theta \, d\varphi ,
\label{Nparaleloang}
\end{eqnarray}
while perpendicular departures are endowed with the probabilities
\begin{eqnarray}
dP_{fi}\;&=&\; \frac{g^2}{2 \,(2\pi)^2 \,m\, \Omega} \,k^3\;\sin \theta \, \cos^2\theta \, \sin^2(k a \cos\theta) \nonumber \\ &\times & |\widetilde{y}_\perp(\Omega + k )|^2 \, dk\, d\theta \\
\;&\equiv &\; \frac{g^2}{2 \,(2\pi)^2 \,m\, \Omega} 
\, p_\perp(ka,\theta)
|\widetilde{y}_\perp(\Omega + k )|^2 \,k^3\sin\theta dk\, d\theta\, \nonumber .
\label{Nperpang}
\end{eqnarray}
In Figs. \ref{Nparalelo} and \ref{Nperp}  we plot the function $p_\shortparallel^N$ and $p_\perp^N$, respectively. Although less evident 
than in the Dirichlet case, $p_\perp^N$ shows a quadrupole
pattern at small distances, being proportional to
$\cos^4(\theta)$.

Yet again, the total probabilities are consistent with the results obtained in the effective action approach.

For completeness we will study in this section the case of the spontaneous decay process with Neumann boundary conditions on the perfect mirror. 
In the case of parallel to the plane oscillatory motion of the atom, the probability density for such process is obtained 
\begin{eqnarray}
dP_{fi} &=&  \frac{d^3{\mathbf k}}{(2\pi)^3} \,
\frac{g^2}{2 m \, \Omega \,k} \, \cos^2(k_z  a) \; \Big[ 4 \pi^2 T \delta(\Omega - k) \nonumber \\ &+&  
k^a k^b \; \widetilde{y}^a(-(-\Omega + k) ) 
\widetilde{y}^b(-\Omega + k ) \Big] \;.
\end{eqnarray}
For the perpendicular motion we get 
\begin{eqnarray}
dP_{fi} &\;=\; & \frac{d^3{\mathbf k}}{(2\pi)^3} \,
\frac{g^2}{2 m \, \Omega \,k} \,  \Big[ 4 \pi^2 T \delta(\Omega - k)  \cos^2(k_z  a) \nonumber \\ &+&  
k_z^2 \; |\widetilde{y}_\perp(- \Omega + k)|^2 \sin^2(k_z  a) \Big]\;.
\end{eqnarray}
The spectrum of the emitted particles is similar to that of the Dirichlet plane. Note, however, that 
the dependence with the mean distance to the mirror is different, as well as the relative weight between the static and non-static contributions for the case of perpendicular motion.

\section{Conclusions}\label{sec:conc}
In this paper we have studied the phenomena of excitation and spontaneous emission of a moving atom in front of a perfect mirror, in a model where the atom is coupled to a quantum real scalar field, and the perfect conductor boundary conditions have been  mimicked by Dirichlet and Neumann boundary conditions.

Assuming that the atom performs small amplitude motions, we computed the vacuum decay probability to the second order in the coupling constant between the electron and the field. This probability is determined by the imaginary part of the effective action, which is a functional of the trajectory of the atom's center of mass. When the atom is close enough to the mirror, the results for Dirichlet and Neumann boundary conditions admit simple interpretations in terms of the images method.

The physical process behind vacuum  decay is, up to this order, the transition of the atom from the ground state to the first excited state, along with the emission of a photon. Thus,  there is a threshold for the process dictated by energy conservation which is $\Omega_{cm} >\Omega$. 
We have also computed the probability for the decay process as a function of the direction of the emitted particle, and checked that the integration over all directions reproduce the vacuum decay probability. The angular dependence of the probability can also be understood in terms of the images method. In particular, for Dirichlet boundary conditions, the radiation emitted by an atom in parallel motion becomes quadrupolar when the atom is close to the mirror. For Neumann boundary conditions this happens for perpendicular motion.

Finally, we analyzed the spontaneous decay of an oscillating atom in front of a mirror. Both the presence of the mirror and the oscillation of the atom induce corrections to the spontaneous emission in free space. The presence of the mirror modifies the angular dependence of the probability, that turns out to be a function of the atom-mirror distance, but not the energy of the emitted photons. When the atom oscillates, the spectrum of the emitted particles is modified by the appearance of two lateral peaks, symmetric with respect to the central peak that corresponds to the static atom.
These results are similar to those for a static atom in front of an oscillating mirror, and opens new possibilities for the eventual experimental observation of the effect. The extension of our results to the 
more realistic case of the electromagnetic field can be addressed, in the dipole approximation, using similar techniques. Work in this direction is in progress.
\section*{Acknowledgments}
This research was supported by Agencia Nacional de Promoci\'on Cient\'ifica
y Tecnol\'ogica (ANPCyT), Consejo Nacional de Investigaciones Cient\'ificas
y T\'ecnicas (CONICET), Universidad de Buenos Aires (UBA) and Universidad
Nacional de Cuyo (UNCuyo).

\end{document}